\documentstyle[amssymb,preprint,aps]{revtex}

\begin{document}
\title{Relaxation effects in the transition temperature of superconducting HgBa$_{2}
$CuO$_{4+\delta }$}
\author{S. Sadewasser and J. S. Schilling}
\address{Department of Physics, Washington University\\
C.B. 1105, One Brookings Dr., St. Louis, MO 63130\vspace{0.5cm}}
\author{J. L. Wagner}
\address{Department of Physics, University of North Dakota\\
P.O. Box 7129, Grand Forks, ND 58202-7129\vspace{0.5cm}}
\author{O. Chmaissem, J. D. Jorgensen, D. G. Hinks and B. Dabrowski}
\address{Materials Science Division, Argonne National Laboratory\\
9700 S. Cass Avenue, Argonne, IL 60439}
\date{\today}
\maketitle

\begin{abstract}
In previous studies on a number of under- and overdoped high temperature
superconductors, including YBa$_{2}$Cu$_{3}$O$_{7-y}$ and Tl$_{2}$Ba$_{2}$CuO%
$_{6+\delta }$, the transition temperature $T_{c}$ has been found to change
with time in a manner which depends on the sample's detailed temperature and
pressure history. This relaxation behavior in $T_{c}$ is believed to
originate from rearrangements within the oxygen sublattice. In the present
high-pressure studies on HgBa$_{2}$CuO$_{4+\delta }$ to 0.8 GPa we find
clear evidence for weak relaxation effects in strongly under- and overdoped
samples ($T_{c}\simeq $ \ 40 - 50 K) with an activation energy $E_{A}$(1
bar) $\simeq $ 0.8 - 0.9 eV. \ For overdoped HgBa$_{2}$CuO$_{4+\delta }$ $%
E_{A}$ increases under pressure more rapidly than previously observed for YBa%
$_{2}$Cu$_{3}$O$_{6.41}$, yielding an activation volume of +11 $\pm $ 5 cm$%
^{3}$; \ the dependence of $T_{c}$ on pressure is markedly nonlinear, an
anomalous result for high-$T_{c}$ superconductors in the present pressure
range, giving evidence for a change in the electronic and/or structural
properties near 0.4 GPa.
\end{abstract}

\newpage 

\section{Introduction}

The study of the mercury-oxide family HgBa$_{2}$Ca$_{n-1}$Cu$_{n}$O$%
_{2n+2+\delta }$ is of particular interest since these cuprates possess the
highest transition temperatures $T_{c}$ of all known high-temperature
superconductors (HTSC). \ For the Hg-compound containing one CuO$_{2}$%
-layer, HgBa$_{2}$CuO$_{4+\delta }$ (Hg-1201), $T_{c}$ reaches values as
high as 98 K at ambient pressure \cite{first-hg1201,Hg-Klehe} and can
reportedly be varied over the entire range from extreme under- to overdoped
solely by changing the oxygen content \cite{xiong1}. \ The Hg-compound
containing three CuO$_{2}$-layers exhibits the highest ambient-pressure
transition temperature ($T_{c}\simeq $ 134 K) reported to date, increasing
to $T_{c}\approx $ 164 K under 30 GPa pressure \cite{Tc164-Gao}.

High-pressure experiments are potentially an important tool to further our
understanding of the superconducting state \cite{schill,schill-japan,schill9}%
. \ For the Hg-compounds the dependence of $T_{c}$ on pressure was first
studied by Klehe et al. \cite{Hg-Klehe} on optimally doped Hg-1201 where $%
T_{c}$ was found to increase linearly with hydrostatic pressure to 1 GPa at
the rate + 1.75 K/GPa. \ An identical initial pressure dependence was
subsequently observed by Klehe et al. \cite{Hg-Klehe2} for the two- and
three-layer compounds HgBa$_{2}$CaCu$_{2}$O$_{6+\delta }$ (Hg-1212) and HgBa$%
_{2}$Ca$_{2}$Cu$_{3}$O$_{8+\delta }$ (Hg-1223); this parallelism in the $%
T_{c}$(P)-dependence for all three optimally doped Hg-compounds was later
confirmed to much higher pressures (%
\mbox{$>$}%
30 GPa) by Gao et al. \cite{Tc164-Gao} who observed that $T_{c}$ initially
increases, but passes through a maximum and decreases at higher pressure. \
In fact, for the majority of HTSC compounds near optimal doping, $T_{c}$(P)
is found to follow a similar bell-shaped curve \cite{schill,takahashi}.

Cao et al. \cite{pressure-on-Hg} studied the pressure dependence of $T_{c}$
for Hg-1201 to 1.5 GPa at widely differing doping levels by varying the
oxygen concentration; Qiu et al. \cite{qiu1} extended these studies to 18
GPa for selected samples with $T_{c}$ values at or above 80 K. \ Jansen and
Block \cite{jansen1} have attempted a quantitative analysis of these
experiments using an indirect-exchange pairing mechanism between conduction
electrons via oxygen anions; the doped oxygen thus fulfills a dual role as
both hole dopant and Cooper pair mediator. \ In this model calculation
pressure effects on $T_{c}$ are assumed to originate solely from a change in
unit cell volume; possible rearrangements of oxygen defects within the unit
cell when pressure is applied are not considered.

Before attempting to gain information about the superconducting state from a
detailed analysis of $T_{c}$(P)-data on the superconducting oxides, it is
essential to know, on a microscopic scale, what structural changes actually
occur under pressure. \ It would be hoped that $T_{c}(P)$ is indeed a
reversible, single-valued function of pressure, as in conventional
superconductors. \ This has appeared to be the case in the vast majority of
experiments on the superconducting oxides where the pressure was changed at
ambient temperature. \ Surprisingly, experiments carried out by Sieburger et
al. \cite{sieburger-tl2201} on Tl$_{2}$Ba$_{2}$CuO$_{6+\delta }$ (Tl-2201)
revealed that the initial dependence of $T_{c}$ on pressure can change by an
order of magnitude, depending on the temperature at which the pressure is
varied; even the sign of $dT_{c}/dP$ may change. \ $T_{c}$(P) is thus {\em %
not} a unique function of pressure but depends strongly on the detailed
temperature-pressure history of the sample. \ Because this anomalous
behavior is a sensitive function of the concentration of interstitial oxygen 
$\delta $ in Tl-2201, it has been attributed \cite{sieburger-tl2201} to
relaxation effects within the oxygen sublattice; indeed, evidence for strong
relaxation effects in $T_{c}$ from oxygen ordering were first obtained by
Veal et al. \cite{Vealquench} in temperature-quench experiments on strongly
underdoped YBa$_{2}$Cu$_{3}$O$_{7-y}$ (Y-123). \ Evidence for relaxation
effects in Y-123 from high-pressure data was obtained somewhat later \cite
{fietz1,fietzdoping}.

In the meantime many oxide superconductors have been found which exhibit
pressure-induced relaxation effects \cite{schill-japan}, including TlSr$_{2}$%
CaCu$_{2}$O$_{7-\delta }$ (Tl-1212) \cite{Tl-1212}, Nd-123 \cite{r'10},
Gd-123 \cite{r'11}, superoxygenated La$_{2}$CuO$_{4+y}$ \cite{r'12}, and
possibly Sr$_{2}$CuO$_{2}$F$_{2+y}$ \cite{r'13}. \ For all these systems,
except Tl-2201, relaxation effects become activated for temperatures above
200 K. \ In Tl-2201 two distinct relaxation processes are activated in the
temperature ranges 15 - 100 K and 200 - 300 K, respectively \cite
{klehetl,Looney-Tl}. \ In view of the importance of relaxation phenomena in
many high-T$_{c}$ superconductors, it would be of interest, before carrying
out detailed comparisons between theory and experiment, to determine whether
or not they play a role in the Hg-oxide system where the values of $T_{c}(P)$
are relatively high, and, at least in Hg-1201, the oxygen defect structure
is known to be quite complex \cite{Jorgensen-defects}. \ In Hg-1212 and
Hg-1223 evidence for oxygen ordering along the c-axis has been obtained in
energy dispersive synchrotron X-ray diffraction experiments for pressures
above $\sim $ 15 GPa \cite{gatt1}; the effect of this ordering on $T_{c}$
was not studied.

In this paper we show evidence for the existence of weak pressure-induced
relaxation effects in $T_{c}$ for both under- and overdoped Hg-1201\ which
are activated at temperatures above 200 K. \ Previous $T_{c}(P)$ studies by
us on optimally doped Hg-1201 \cite{Hg-Klehe} gave no evidence for
relaxation behavior. \ In the overdoped Hg-1201 samples the pressure
dependence $T_{c}(P)$ to 0.8 GPa is clearly nonlinear, providing evidence
for a possible pressure-induced electronic and/or structural modification.

\section{Experiment}

In the present experiments one underdoped sample and two overdoped samples
(A and B) of polycrystalline HgBa$_{2}$CuO$_{4+\delta }$ with transition
temperatures in the range 40 - 50 K were studied. \ Typical sample sizes
were 2$\times $1$\times $1 mm$^{3}$ with masses ranging from 4 to 12 mg. \
All samples were synthesized in an evacuated quartz tube as described
previously \cite{Cu-subst-on-Hg}. \ Both the single underdoped sample and
the overdoped sample A were obtained from a single synthesis by subsequent
annealing to produce samples with significantly different oxygen contents
but similar values of $T_{c}$. \ The overdoped sample A ($T_{c}$(1 bar) $%
\simeq $ 47.81 K) was annealed in 150 atm. oxygen at 220$^{\circ }C$ for 100
h followed by a slow cool to room temperature, whereas the underdoped sample
($T_{c}$(1 bar) $\simeq $ 47.55 K) was annealed in flowing argon at 400$%
^{\circ }C$ for 24 h and then rapidly cooled to room temperature. \ The
overdoped sample B ($T_{c}$(1 bar) $\simeq $ 40.0 K) was obtained from a
second synthesis in powder form and annealed in 30 atm. oxygen at 200$%
^{\circ }C$ for 100 h. \ Neutron diffraction experiments on these samples
determined the oxygen contents at x $\approx $ 0.22 \cite{substamount} and
0.12 \cite{Jorgensen-defects} for the overdoped sample A and the underdoped
sample, respectively.

The He-gas pressure system (Harwood) used in this study is capable of
generating hydrostatic pressures to 1.4 GPa. The CuBe pressure cell
(Unipress) is inserted into a two-stage closed-cycle refrigerator (Leybold)
which operates in the temperature range 2 - 320 K. The pressure in the cell
can be changed at any temperature above the melting curve T$_{m}$ of the He
pressure medium \cite{melting} (for example, $T_{m}\simeq $ 13.8 K at 0.1
GPa and $T_{m}\simeq $ 52.4 K at 0.8 GPa). The decrease in pressure upon
cooling is kept to below 20\% by the introduction of a sizeable room
temperature ``dead volume''. Unless otherwise stated, the value of the
pressure given was determined at a temperature near $T_{c}$.

The superconducting transition of the Hg-1201 samples is measured by the ac
susceptibility technique using a miniature primary/secondary coil system
located inside the 7 mm I.D. bore of the pressure cell. In the present
studies a magnetic field of 1.1 Oe (rms) at 507 Hz is applied along the long
dimension of the sample. Further details of the experimental setup are given
elsewhere \cite{expdetails}.

\section{Results of Experiment}

\subsection{Overdoped Sample}

Fig. 1 shows the transition to superconductivity for the overdoped Hg-1201
sample A at ambient pressure. \ The transition is measured by sweeping
through the temperature range 10 K\ to 55 K over a 2 h period, yielding the
value $T_{c}$(1 bar) $\simeq $ 47.81 K (inset in Fig. 1) for the
superconducting onset. \ The size of the transition corresponds to full
shielding, if the demagnetization factor ${\cal D}$ is neglected \cite
{comment-density}. \ The transition temperature in the subsequent
measurements under high pressure is determined by shifting the upper 15\% of
the full transition along the temperature axis until overlap is achieved
with the initial measurement. \ Only the upper portion of the transition is
used because this region is independent of the applied magnetic field.
Through this procedure an accuracy of $\pm $ 20 mK\ for the relative $T_{c}$
shifts under pressure could be achieved. \ In most experiments the
transition was remeasured for verification.

The first set of high-pressure data for the overdoped Hg-1201 sample A (open
circles in Fig. 2a) was taken by applying 0.16 GPa at 298 K and holding this
pressure for one hour, followed by a rapid cooldown (3 - 4 K/min) to measure 
$T_{c}$. At temperatures near $T_{c}$ the measured pressure had decreased to
0.12 GPa due to the thermal contraction of the helium pressure medium upon
cooling. \ The sample was then heated back to 298 K, the pressure increased
and $T_{c}$ remeasured. \ This procedure was repeated up to the highest
pressure of 0.61 GPa (0.54 GPa at 50 K). \ The pressure dependence $T_{c}(P)$
is seen to be markedly nonlinear, an anomalous result in this pressure range
for high-T$_{c}$ oxides.

The second set of data on this sample (solid squares in Fig. 2a) was taken
by applying 0.79 GPa at 298 K (0.71 GPa at 50 K) followed by a one hour
anneal at this temperature. \ After measuring $T_{c}$ and heating back up to
298 K, the pressure was decreased by $\sim $ 20\%, held for one hour at 298
K, and $T_{c}$ remeasured. This procedure was then repeated several times
until the pressure was fully released; $T_{c}$ is seen to return to its
initial value at 47.8 K. \ For pressure changes at room temperature (RT), $%
T_{c}(P)$ thus appears to be a reversible, but markedly nonlinear, function
of pressure with initial slope $(dT_{c}/dP)_{RT}\simeq $ -1.55 $\pm $ 0.26
K/GPa.

We now determine the hydrostatic pressure dependence of $T_{c}$ for pressure
changes at low temperatures, an experiment that can best be carried out
using a He-gas high-pressure system. \ In the third set of data on the
overdoped sample A (solid diamonds in Fig. 2a), we first apply and hold 0.76
GPa pressure at 298 K for 10 min before cooling down rapidly to measure $%
T_{c}$ ($P=$ 0.69 GPa at 50 K). \ Some pressure was then added at 60 K to
obtain 0.75 GPa and $T_{c}$ was remeasured. \ The pressure was then
successively decreased at low temperatures (T 
\mbox{$<$}%
55 K) and $T_{c} $ redetermined. \ The value of $T_{c}$ at ambient pressure
is now seen to lie 0.26 K below its initial value. Only after annealing the
sample at 298 K for 40 min does $T_{c}$ shift back to its initial value
(47.81 K) within the experimental accuracy of $\pm $ 0.02 K. \ These data
demonstrate that the pressure dependence of $T_{c}$ is clearly nonlinear.
The fact that $T_{c}(P)$ depends on the temperature at which the pressure is
changed gives evidence for relaxation behavior in overdoped Hg-1201, as
discussed in the Introduction; the time dependence of this relaxation will
be examined below. \ From Fig. 2a the initial slope near 1 bar for pressure
changes at low temperature is $(dT_{c}/dP)_{LT}\simeq $ -1.15 $\pm $ 0.52
K/GPa.

To establish whether the above nonlinearity in $T_{c}(P)$ is an accidental
occurrence or is characteristic for overdoped Hg-1201 samples, a second
overdoped sample (B) with $T_{c}\simeq $ 40.0 K from a different synthesis
was studied. \ As seen in Fig. 2b, pressure was applied successively at 298
K (open circles) and released in steps (closed circles) at low temperature (%
\mbox{$<$}%
65 K). \ This procedure was then repeated (open and closed triangles) to
check the reproducibility of the data. \ Evidence for nonlinearity in $%
T_{c}(P)$ is also seen in Fig. 2b for this sample; however, the pressure
above which the change in slope occurs appears to be shifted to higher
pressures for sample B ($\sim $ 0.5 GPa) relative to sample A ($\sim $ 0.4
GPa). \ For sample B we find the initial slopes $(dT_{c}/dP)_{RT}\simeq $
-0.66 $\pm $ 0.18 K/GPa and $(dT_{c}/dP)_{LT}\simeq $ -0.57 $\pm $ 0.30
K/GPa which have approximately half the magnitude of those for the less
strongly overdoped sample A. \ The magnitude and even the sign of $%
(dT_{c}/dP)_{RT}$ differ from the results of Cao et al. \cite{pressure-on-Hg}
on similarly overdoped Hg-1201; the reason for this discrepancy is unknown,
but may arise from differences in the pressure medium (He versus
Flourinert), pressure technique, details of sample preparation or
measurement technique (ac susceptibility versus resistivity). \
Interestingly, the data of Cao et al. \cite{pressure-on-Hg} for strongly
overdoped Hg-1201 reveal a nonlinearity in $T_{c}(P)$ near 0.5 GPa, but the
sign of the change in slope $dT_{c}/dP$ is opposite to ours.

We would now like to determine the temperature at which the relaxation
behavior in the overdoped Hg-1201 sample A initiates. \ The sample studied
(second piece from the same pellet) exhibited $T_{c}\simeq $ 47.62 K at
ambient pressure (pt. 1 in Fig. 3). Applying and holding 0.67 GPa pressure
at 298 K for 1 h, $T_{c}$ decreased to pt. 2 (47.18 K). \ Immediately
following the $T_{c}$ measurement, the pressure was fully released at 50 K,
yielding $T_{c}\simeq $ 47.32 K (pt. 3 in Fig. 3), which lies 0.3 K lower
than the initial value at ambient pressure. \ The value of $T_{c}$ is found
not to change following anneals for 1 h at 100 K\ (pt. 4) and then for
another hour at 200 K (pt. 5). However, $T_{c}$ is seen to relax back toward
its initial value following 1 h (pt. 6) and then 3 h (pt. 7) anneals at 298
K. \ The data in Fig. 3 thus show that relaxation effects are only activated
for temperatures above 200 K. \ We cannot exclude the possibility that
relaxation phenomena occur at temperatures below 50 K since in the present
studies the measurement of $T_{c}$ itself constitutes annealing the sample
at $\sim $ 50 K.

We now examine the time-dependence of the relaxation phenomena for overdoped
Hg-1201 at three different pressures. \ In these measurements a third piece
of sample A was studied with $T_{c}\simeq $ 47.81 K at ambient pressure;
full shielding was exhibited \cite{comment-density}. \ A pressure of 0.61
GPa was applied and held 1 h at 298 K\ before cooling to measure $%
T_{c}\simeq $ 47.38 K. \ After the full release of pressure at 48 K, we find 
$T_{c}\simeq $ 47.58 K (pt. 1 in Fig. 4a). The sample was then heated to 265
K and held for 1 h, before cooling rapidly to measure $T_{c}\simeq $ 47.64 K
(pt. 2 in Fig. 4a). \ Subsequent annealing at 265 K for longer times gives
the $T_{c}$ values represented by points 3 to 7 in Fig. 4a. \ The fit to the
data, which yields a relaxation time $\tau \simeq $ 3.4 h, is discussed
below. \ The difference of 70 mK between the value of $T_{c}$ at the outset
of the experiment (47.81 K)\ and the final value (47.74 K)\ is likely due to
the different annealing temperature (here $T_{a}=$ 265 K versus 298 K for
the initial measurement). \ We previously observed a similar effect in YBa$%
_{2}$Cu$_{3}$O$_{6.41}$ \cite{Vealquench,ownYBCO}. \ As we discuss in more
detail below, a lower annealing temperature leads to a higher state of
oxygen order which increases the\ hole-carrier concentration in the CuO$_{2}$
planes, shifting $T_{c}$ up for an underdoped sample or down for an
overdoped sample, as observed.

Our previous studies on YBa$_{2}$Cu$_{3}$O$_{6.41}$ \cite{ownYBCO} revealed
that the relaxation behavior itself is strongly pressure dependent. To
address this question for the present overdoped Hg-1201 sample A, a pressure
of 0.62 GPa was applied and held at 298 K for 1 h. \ Following the partial
release of pressure to 0.13 GPa at 53 K, $T_{c}$ was found to be 47.41 K
(pt. 1 in Fig. 4b). \ Subsequent annealing at $T_{a}\simeq $ 265 K (P $%
\simeq $ 0.15 GPa at this temperature) for various time periods up to a
total of 145 h gives the relaxation behavior represented by points 2 to 8 in
Fig. 4b. A cursory comparison of the data in Figs. 4a and 4b reveals that
the relaxation time $\tau ($265 K) has become longer under pressure. \ To
determine $\tau $ at higher pressures, the sample was then subjected to 0.7
GPa pressure at 298 K for 1.6 h which was then partially released to 0.25
GPa at 52 K (pt. 1 in Fig. 4c). \ Because of the rapid increase in $\tau $
with pressure, the annealing temperature was increased to 285 K to speed up
the relaxation. The sample was annealed up to a total time of 42.5 h (pts. 2
to 6 in Fig. 4c).

We now analyze the time-dependent relaxation behavior of $T_{c}$ in Fig. 4
in terms of the stretched exponential \cite{r9,r10} 
\begin{equation}
T_{c}(t)|_{_{relax}}=T_{c}(\infty )-[T_{c}(\infty )-T_{c}(0)]\exp \left\{
-\left( \frac{t}{\tau }\right) ^{^{\alpha }}\right\} ,
\end{equation}
where $T_{c}(\infty )$ is the transition temperature at a given pressure
after infinite time, $T_{c}(0)$ is the initial value of the transition
temperature at a given pressure before relaxation begins, $\tau $ is the\
temperature- and pressure-dependent relaxation time, and $\alpha \leq 1$ is
the stretched exponent. \ In a previous analysis of relaxation behavior on a
strongly underdoped Y-123 single crystal \cite{ownYBCO}, where the
pressure-induced shifts in $T_{c}$ were large and the data could be analyzed
to high accuracy, we found $\alpha $ = 0.6, independent of the pressure. \
For the present overdoped Hg-1201 compound the measured shifts in $T_{c}$
are much smaller, resulting in relatively large error bars. \ In the
following we use $\alpha $ = 0.6 \cite{comment-on-alpha} and set the
transition temperature at zero time, $T_{c}(0)$, to the value measured
immediately after the pressure release at low temperatures, before any
relaxation could take place.

A fit with Eq. (1)\ to the ambient-pressure relaxation data at 265 K in Fig.
4a (solid line) is seen to describe the data well within experimental error,
yielding the parameters $T_{c}(\infty )\simeq (47.74\pm 0.01)$ K and $\tau ($%
265 K$)\simeq (3.4\pm 0.6)$ h. \ A second experiment (not shown) under
similar conditions gives $\tau ($265 K$)\simeq (3.9\pm 0.7)$ h, in agreement
with the previous value. \ A fit to the relaxation data at 0.15 GPa and 265
K in Fig. 4b yields $T_{c}(\infty )\simeq (47.55\pm 0.01)$ K and $\tau ($265
K$)\simeq (17.5\pm 5.4)$ h. \ This relaxation time is distinctively longer
than the ambient pressure values, implying that the relaxation slows down
rapidly under pressure. \ Since the relaxation at 0.30 GPa and 285 K in Fig.
4c proceeded somewhat more rapidly than anticipated, a small correction to
the annealing time is necessary to account for the finite time spent
warming(cooling) the sample to(from) the annealing temperature. \ Based on
the relaxation time obtained from a preliminary fit to the data with Eq.
(1), we estimate that this correction amounts to only $\sim $ 4 min. \ The
corrected data in Fig. 4c are now fit with Eq. (1), yielding $T_{c}(\infty
)\simeq (47.46\pm 0.01)$ K and a relaxation time $\tau ($285 K$)\simeq
(0.92\pm 0.10)$ h.

\subsection{Underdoped Sample}

We now study the pressure dependence of $T_{c}$ on an underdoped Hg-1201
sample with $T_{c}$(1 bar) $\simeq $ 47.55 K, an almost identical value to
that for the overdoped sample A. \ Since the superconducting onset is less
sharp than that for the overdoped sample, changes in $T_{c}$ under pressure
cannot be determined as accurately ($\pm $ 50 mK versus $\pm $ 20 mK).
Following the initial measurement of $T_{c}$ at ambient pressure (pt. 1 in
Fig. 5), a pressure of 0.79 GPa was applied at 298 K\ and held for 15
minutes before cooling down to measure $T_{c}\simeq $ 48.55 K (pt. 2 in Fig.
5); the pressure at this temperature was 0.71 GPa, yielding the pressure
derivative $(dT_{c}/dP)_{RT}\simeq $ +1.41 $\pm $ 0.14 K/GPa, a value
somewhat less than that (+2 K/GPa) reported for underdoped Hg-1201 by Cao et
al. \cite{pressure-on-Hg}. \ Pressure was then successively decreased at
temperatures below 90 K and $T_{c}$ immediately remeasured each time (pts. 3
to 9 in Fig. 5), revealing a linear dependence of $T_{c}$ on pressure $%
(dT_{c}/dP)_{LT}\simeq $ +1.15 $\pm $ 0.05 K/GPa. \ After the pressure was
released (pt. 9), $T_{c}$ was found to lie 150 mK higher than the initial
value (pt. 1). \ Annealing the sample at 298 K for a total of 4 h (pts. 10 -
12) fully restored the initial $T_{c}$ value.

Annealing this sample for extended time periods at RT apparently causes some
sample degradation, as evidenced by irreversible changes in the shape of the
superconducting transition. \ A strong degradation of underdoped Hg-1201
samples stored in air was recently reported by Peacock et al. \cite{peacock}
in which $T_{c}$ increased from 30 K to 75 K over the time span of 6.5 weeks
following their synthesis. \ This gradual shape change in the transition and
the smallness of the relaxation effect itself precluded a detailed
measurement of the time, temperature and pressure dependence of the
relaxation phenomena for this underdoped sample; from the data, however, we
can give a rough estimate of the relaxation time at ambient pressure, namely 
$\tau ($298 K) $\approx $ 1 h.

\section{Discussion}

\subsection{Intrinsic Pressure Dependence of $T_{c}$}

Pressure-induced relaxation effects have been observed in several high-$T_{c}
$ superconductors, including Y-123, Tl-2201 and Tl-1212 \cite{schill-japan}.
\ Very recent studies on underdoped Y-123 have shown that $T_{c}(P)$ to 17
GPa depends markedly on whether the pressure is applied at ambient or low
temperatures \cite{tissen1,sascha4,saschathesis}. \ In oxide
superconductors, therefore, the measured dependence of $T_{c}$ on pressure
may depend in a sensitive manner on the detailed pressure/thermal history of
the sample. \ The high pressure environment does not {\it create} the
relaxation processes, but rather makes those already present under ambient
conditions more readily visible than in conventional temperature-quench
experiments.

Theoretical calculations of the properties of matter, including their
superconductivity, are generally carried out for a given fixed set of
lattice parameters and atomic coordinates; pressure effects are generally
estimated under the assumption that the change in properties, like $T_{c},$
is a function of the anisotropic decrease in the lattice parameters, with
all atomic separations scaling accordingly. \ Before a measured $T_{c}(P)$%
-dependence can be compared in a quantitative way with theory, therefore, it
should be purged of all influences from relaxation behavior. \ Since most
high-pressure cells only allow the pressure to be changed at room
temperature, many results from the large body of high-pressure studies on
high-$T_{c}$ materials do contain relaxation effects and thus must be
repeated using high-pressure technology similar to ours to separate
intrinsic from relaxation effects.

In the oxide superconductors the measured value of the transition
temperature $T_{c}^{meas}(\delta ,P,T_{a},t)$ is a function of both the
doping level $\delta $ and the sample's detailed pressure/temperature
history (i.e. the length of time $t$ the sample is annealed at a given
temperature $T_{a}$ following a change in pressure $P$); it can be
represented by the expression 
\begin{equation}
T_{c}^{meas}(\delta ,P,T_{a},t)=T_{c}^{intrin}(\delta ,P)+\Delta
T_{c}^{relax}(\delta ,P,T_{a},t).
\end{equation}
The first term on the right side gives the ideal or ``intrinsic'' change in $%
T_{c}$ which occurs immediately following a change in pressure at any
temperature, whereas the second ``relaxation'' term takes into account the
change in $T_{c}$ from pressure-, temperature- and time-dependent relaxation
effects. \ The simplest way to determine $T_{c}^{intrin}(P),$ the dependence
which can be compared most directly with theory, is to carry out the entire
experiment at a temperature low enough to eliminate (freeze out) all
relaxation processes, so that $\Delta T_{c}^{relax}=$ 0. \ As seen in Fig.
3, for Hg-1201 relaxation processes appear to be frozen out for $T\leq $ 200
K. \ The two distinct ``intrinsic'' and ``relaxation'' contributions to $%
T_{c}^{meas}(P,T_{a},t)$ for underdoped Hg-1201 can be easily recognized in
Fig. 5 where the application of 0.79 GPa pressure at RT is seen to cause a
larger change in $T_{c}$ than the pressure release below 90 K, corresponding
to the pressure derivatives $(dT_{c}/dP)_{RT}\simeq (+1.41\pm 0.14)$ K/GPa
and $(dT_{c}/dP)_{LT}=(dT_{c}^{intrin}/dP)\simeq (+1.15\pm 0.05)$ K/GPa,
respectively.

Whereas in the {\it underdoped} sample the change in $T_{c}$ with pressure
is linear, in Figs. 2a and 2b it is seen that for {\it overdoped} Hg-1201
there is a distinct change in the slope of $T_{c}(P)$ near 0.4 GPa for
sample A and near 0.5 GPa for sample B. \ This evident nonlinearity is
retained in the ``intrinsic'' $T_{c}^{intrin}(P)$ dependence obtained by
releasing the pressure at low temperatures. This gives evidence that this
nonlinearity is an intrinsic effect and does not have its origin in
relaxation effects which occur at temperatures above 55 K, but may be due to
a structural change. \ As pointed out previously, Cao et al. \cite
{pressure-on-Hg} also observe a nonlinearity in $T_{c}(P)$ near 0.5 GPa in
strongly overdoped Hg-1201, but the change in slope is opposite in sign to
the present results. \ Chen et al. \cite{chen1} report a negative slope
change in $T_{c}(P)$ near 0.5 GPa in nearly optimally doped Hg-1212. \
Recent neutron powder diffraction experiments under high pressure on Hg-1201
at variable oxygen content find no evidence for a structural phase
transition to 1 GPa \cite{aksenov1}; this study, however, did not include
heavily overdoped samples with $T_{c}<$ 80 K, as in both the present
experiments and those of Cao et al. \cite{pressure-on-Hg}. \ Further
high-pressure structural studies are required to clarify the origin of the
nonlinearity in $T_{c}(P)$.

The nonlinear $T_{c}(P)$ dependence and the notable change in $dT_{c}/dP$ in
the present experiments on the two heavily overdoped Hg-1201 samples are but
two of a number of indications that sufficiently overdoped Hg-1201 (where $%
T_{c}\leq $ 80 K) possesses basic (structural) differences to underdoped
Hg-1201. \ Cao et al. \cite{pressure-on-Hg} have shown that the initial
pressure derivative $dT_{c}/dP$ changes radically in the overdoped regime
for $T_{c}\leq $ 80 K. \ Xiong et al. \cite{xiong1} note that the $c$
lattice constant increases monotonically with oxygen content over the entire
underdoped and partially overdoped regime, until $T_{c}\approx $ 80 K is
reached, remaining constant thereafter; the valence-bond sum technique for
estimating carrier concentration also gives anomalous results for an oxygen
content above this critical value. \ Xiong et al. \cite{xiong1} suggest that
there may be two different types of lattice sites available for oxygen
defects. \ In fact, recent powder neutron diffraction experiments on Hg-1201
by Jorgensen et al. \cite{Jorgensen-defects} over a wide range of oxygen
defect concentrations suggest that there are two competing oxygen defects in
the Hg layer, an O3 defect in the center of the unit cell and an O4 defect
nearer the cell's edge. \ The relative concentrations of these defects
switches with increasing $\delta $ upon passing from the underdoped region
and through the maximum $T_{c}$, where O4 is the dominant defect, to the
overdoped region where, at $T_{c}\simeq $ 80 K, O3 suddenly becomes the
dominant defect. \ This ``change of the defect'' is mirrored in the
dependence of $T_{c}$ on $\delta $ which deviates from the canonical
inverted parabolic behavior. \ Of particular relevance to the present
experiments is the fact that a sudden {\it increase }in the unit cell volume
by $\sim $ 1\% accompanies this ``change of the defect'' O4 $\rightarrow $\
O3 at $T_{c}\simeq $ 80 K \cite{Jorgensen-defects}. \ Since O3 should be the
sole defect in our strongly overdoped samples, this implies that if
sufficiently high pressure is applied, the O4 defect should become
energetically favored over O3. \ In this picture the nonlinearity in $%
T_{c}(P)$ would signal a sudden changeover from a dominant O3 defect to O4.
\ The fact that the critical onset pressure for sample B is somewhat larger
(0.5 GPa) than for sample A (0.4 GPa) supports this picture since sample B
is more strongly overdoped and thus the O3 defect more stable. \ Neutron
diffraction experiments under high pressure conditions would be desirable to
clarify the anomalous but interesting overdoped state in Hg-1201.

It is interesting to note that for under- and overdoped Hg-1201 with $%
T_{c}\approx $ 49 K the initial ``intrinsic'' slopes $%
dT_{c}^{intrin}/dP=(dT_{c}/dP)_{LT}\simeq $ -1.15 $\pm $ 0.05 K/GPa and
-1.15 $\pm $ 0.52 K/GPa, respectively, are equal in magnitude but opposite
in sign. \ At first glance this result would appear to support a simple
charge-transfer model \cite{schill} in which $%
dT_{c}^{intrin}/dP=(dT_{c}/dn)(dn/dP)$ and $dn/dP=$ constant 
\mbox{$>$}%
0. \ However, this would lead to the expectation that $dT_{c}^{intrin}/dP%
\simeq $ 0 for an optimally doped sample (where $dT_{c}/dn=$ 0), contrary to
the experimental result that $dT_{c}^{intrin}/dP\simeq $ +1.75 $\pm $ 0.1
K/GPa \cite{Hg-Klehe}. \ We note that the intrinsic pressure dependence for
the more strongly overdoped sample B is only half the magnitude of that for
sample A; the charge transfer model would lead to the expectation that $%
dT_{c}^{intrin}/dP$ should be {\it greater} in magnitude, contrary to
experiment. \ It is certainly possible that pressure-dependent shifts in the
defect structure in Hg-1201 play a role here. \ Further measurements of $%
T_{c}(P)$ together with systematic structural studies under high pressure
are needed to resolve this issue.

\subsection{Relaxation Behavior at Ambient Pressure}

In order to better characterize the relaxation behavior at different
pressures and temperatures in overdoped Hg-1201, we calculate the
pressure-dependent activation energy $E_{A}(P)$ from the measured values of
the relaxation time $\tau (T,P)$ at a given temperature using the Arrhenius
law

\begin{equation}
\tau (T,P)=\tau _{0}\cdot exp\left\{ \frac{E_{A}(P)}{k_{B}T}\right\} .
\end{equation}
We set the attempt period $\tau _{o}$ equal to the value ($\tau _{0}\approx $
1.4$\times $10$^{-12}$ s) found in earlier temperature-quench experiments on
Y-123 \cite{Vealquench}. \ Using the values of the relaxation times at
ambient pressure from the previous section, the corresponding activation
energy for overdoped (sample A) and underdoped Hg-1201 can be calculated
using Eq. 3, yielding $E_{A}($1 bar) $\simeq $ 0.84 eV and 0.90 eV,
respectively. \ These values are quite comparable to each other and to
activation energies determined from relaxation studies on Y-123 (0.97 eV) 
\cite{Vealquench,fietzdoping}, Tl-2201 (0.72 eV) \cite{Looney-Tl} and
Tl-1212 (0.86 eV) \cite{Tl-1212}, as well as to oxygen tracer diffusion
studies on Y-123 and other oxide superconductors \cite{tracerdiffusion}. \
This fact, and the known sensitivity of the relaxation behavior to the
oxygen content \cite{sieburger-tl2201,fietzdoping}, make it likely that the
above relaxation effects originate in the oxygen sublattice. \ To our
knowledge, no attempt has yet been made to compute the energy barrier for
oxygen diffusion in Hg-1201; in a recent model calculation Islam and Winch 
\cite{computer simulation} have estimated $E_{A}($1 bar) $\approx $ 0.68 eV
for the migration of interstitial O(4) oxygen in the three-layer compound
Hg-1223, a value roughly comparable to our present result for Hg-1201.

It is interesting to note that no relaxation whatsoever has been observed in 
$T_{c}$ for optimally doped Hg-1201 \cite{Hg-Klehe} or, for that matter, for
any other optimally doped oxide superconductor studied to date \cite
{Hg-Klehe2,sieburger-tl2201,fietzdoping,Tl-1212}. \ Does this mean that in
optimally doped samples relaxation phenomena do not occur? \ Certainly not!
\ The absence of relaxation phenomena in $T_{c}$ for optimally doped samples
can be readily understood if the change in $T_{c}$ due to relaxation
phenomena originates primarily from a change in the carrier concentration $n$
in the CuO$_{2}$ plane(s); Hall effect studies on Tl-2201 support this
conclusion \cite{taka1}. \ In an optimally doped sample the value of $T_{c}$
is maximized as a function of $n,$ i.e. $T_{c}(n)$ is at an extremum where $%
(dT_{c}/dn)=$ 0, so that $(dT_{c}/dP)^{relax}\simeq
(dT_{c}/dn)(dn/dP)^{relax}\approx $ 0. \ In optimally doped samples,
therefore, relaxation processes in the oxygen sublattice initially cause no
change in $T_{c},$ but should have a measurable effect on other properties
such as the electrical resistivity, Hall coefficient, lattice parameters,
etc. \ In this regard further studies on the Hg-oxides would be of
particular interest since optimal doping occurs at a {\it finite }%
concentration of oxygen defects, in contrast to Y-123 and Tl-1201 which are
nearly devoid of oxygen defects when $T_{c}$ is at its maximum value.

An important question remaining is to identify the mechanism by which the
relaxation processes change the carrier concentration in Hg-1201. \ For
Y-123 Veal and Paulikas \cite{oxordmodel} proposed that the local
rearrangement of oxygen anions in the CuO chain layer causes the valence of
the ambivalent Cu$^{1+/2+}$ cations in this layer to change. \ The
maintainance of overall charge neutrality then dictates that the charge
carrier concentration in the CuO$_{2}$ layers must adjust accordingly. \ In
Tl-2201 the oxygen defects are incorporated interstitially in the Tl$_{2}$O$%
_{2}$ double-layer \cite{ox-in-tl}; some form of local ordering of these
oxygen would then prompt some of the ambivalent Tl$^{3+}$-cations in this
layer to decrease their valence by drawing electron charge away from the CuO$%
_{2}$ plane (for thallium +1 and +3 are the preferred valence states). \ The
presence of ambivalent cations would not appear to be mandatory for charge
transfer to occur, but probably serves to enhance it; as pointed out in the
Introduction, relaxation phenomena have been reported in superoxygenated La$%
_{2}$CuO$_{4+y}$ \cite{r'12}, although this compound contains no ambivalent
cations outside the CuO$_{2}$ layer. \ Hg-1201 would appear to be a further
such example, since XPS studies \cite{Hg-valence} on an optimally doped
sample indicate that the valence of Hg is 2+. \ Further evidence against Hg
cation ambivalence in Hg-oxides is given by chemical analysis on a series of
Hg-1212 samples near optimal doping which show that the average Cu valence
increases with oxygen content, but that the valence of Hg is fixed at 2+ 
\cite{tsuchiya1}. \ Two possible mechanisms for the observed weak relaxation
phenomena in Hg-1201 are: \ (1) the degree of oxygen ordering may directly
influence how the available hole carriers are distributed between the CuO$%
_{2}$ planes and other structural elements \cite{r'12}; (2) ambivalent Cu$%
^{1+/2+}$ cations on Hg sites may switch their valence as oxygen defects
rearrange. \ Indeed, studies by Wagner et al. \cite{Cu-subst-on-Hg} show
that there is a substitutional defect in Hg-1201 in the form of a partial
substitution of Cu ions on the Hg site; the same overdoped Hg-1201 sample A
used in the present experiments was found to contain a substitution of $\sim 
$ 8\% of Cu on the Hg site \cite{substamount} which could conceivably lead
to modest relaxation effects of the order $\sim $ 8\% of that expected for
full substitution. \ It is interesting to note that in the present
experiment on overdoped Hg-1201 (sample A) the magnitude of the observed
relaxation derivative ($(dT_{c}/dP)^{relax}\approx $ -0.3 K/GPa) is only $%
\sim $ 10\% of that for Y-123 \cite{fietzdoping} with a similar value of $%
T_{c}$. \ A critical test would be to study non optimally doped Hg-1201
samples devoid of Cu substitution; if the second scenario is correct, there
would then be no relaxation in $T_{c}$.

\subsection{Relaxation Behavior under High Pressure}

Using the values of $\tau (T,P)$ for overdoped Hg-1201 (sample A) from the
previous section, $E_{A}(P)$ is calculated using Eq. 3 and displayed in Fig.
6. \ We ignore any change in $\tau _{o}$ with pressure since its anticipated
decrease under pressure is very small \cite{note1}. \ In this figure the
value of the pressure near the annealing temperature $T_{a},$ rather than
the value near $T_{c},$ is used since below 250 K only insignificant
relaxation occurs. Assuming that, as found previously for Y-123 \cite
{ownYBCO}, $E_{A}$ is a linear function of pressure, we obtain $%
dE_{A}/dP\simeq $ (+114 $\pm $ 50) meV/GPa (dashed line in Fig. 6). To our
knowledge, the only other such study on a superconducting oxide was on
strongly underdoped YBa$_{2}$Cu$_{3}$O$_{6.41}$ \cite{ownYBCO} where we
found $dE_{A}/dP\simeq $ (+44 $\pm $ 2) meV/GPa, approximately half the
present value for Hg-1201.

The pressure dependence of relaxation behavior is normally analyzed in terms
of an activation volume $\Delta V_{A}$ which can be calculated from the
pressure dependence of the activation energy \cite{activationvolume} 
\begin{equation}
\Delta V_{A}=\Delta V_{f}+\Delta V_{m}=N_{A}\left( \frac{\partial E_{A}}{%
\partial P}\right) ,
\end{equation}
where $\Delta V_{f}$ is the change in volume due to the formation of a
vacancy or interstitial, $\Delta V_{m}$ is the volume change due to the
migration of the diffusing atom, and $N_{A}$ is Avogadro's number. If we
assume that the oxygen defects involved in the relaxation phenomena are
already present and need not be created, then $\Delta V_{f}$ = 0. \ As
discussed above, two distinct types of oxygen defect have been found in the
HgO layer in Hg-1201 \cite{substamount,Jorgensen-defects}, with a transition
from one type to the other near the optimal doping concentration. \ The
nonlinearity in $T_{c}(P)$ seen in Fig. 2 near 0.4 GPa for the overdoped
samples may signal a transition from the O3 to O4 defect type. \ Since in
the present experiments the time dependence of the relaxation behavior was
only studied for pressures below 0.3 GPa, we assume that only a single
oxygen defect type is responsible for the relaxation, so that $\Delta
V_{f}=0 $. \ From Eq. (4) we now estimate from the above value of $dE_{A}/dP$
the molar activation volume to be $\Delta V_{A}$ $\approx \Delta V_{m}\simeq 
$ + (11.0 $\pm $ 4.8) cm$^{3}$/mol which is approximately twice the molar
volume of the diffusing O$^{2-}$ ion (5.78 cm$^{3}$/mol \cite{crc}). \ In
our recent study on underdoped YBa$_{2}$Cu$_{3}$O$_{6.41}$ an activation
volume of + 4.2 cm$^{3}$/mol \cite{ownYBCO} was obtained.

A positive migration volume $\Delta V_{m}$ implies that the lattice
increases in volume as a diffusing ion passes over a saddle point in moving
from one site to another. \ It is physically reasonable that diffusion slows
under pressure ($dE_{A}/dP>$ 0) because squeezing the lattice together
leaves less space through which the interstitial atoms can diffuse. \
Indeed, in ionic compounds \cite{diffusionbook}, such as the oxide
superconductors, a positive activation volume is normally observed. \
However, an enhancement of the diffusion rate under pressure (negative
activation volume) may be observed in covalently bonded substances, such as
amorphous Si \cite{Theisspaper}. \ Recently, the activation volume for both
Hg-1201 and Y-123 has been calculated \cite{saschathesis} using a hard
sphere model to estimate the local expansion in the lattice as the oxygen
defect travels along its diffusion path. \ For both systems the measured
values of $\Delta V_{A}$ are underestimated by a factor of two to three in
this model.

In summary, the present experiments on Hg-1201 give evidence for weak
relaxation phenomena in the transition temperature for both under- and
overdoped samples; these relaxation effects are opposite in sign but roughly
equal in magnitude and involve similar activation energies. This result,
plus the absence of relaxation effects in $T_{c}$ for optimally doped
Hg-1201 \cite{Hg-Klehe}, support the view that relaxation phenomena
influence $T_{c}$ via the transfer of charge in or out of the CuO$_{2}$
planes. The relaxation time at ambient pressure is found to be approximately
4 h at 265 K and increases strongly with pressure. \ This pressure
dependence can be expressed in terms of an activation volume equal to + 11.0
cm$^{3}$/mol. \ An anomalous nonlinearity in $T_{c}(P)$ is observed for the
overdoped samples near 0.4 GPa which may indicate a transition from one
oxygen defect type to another.\vspace{1cm}

\noindent {\bf Acknowledgments}

The work at Washington University (SS and JSS) is supported by the National
Science Foundation under grant DMR 98-03820. Part of this work supported by
the NSF Office of Science and Technology Centers, grant No. DMR 91-20000 (OC
and BD) and US Department of Energy, Basic Energy Sciences - Material
Sciences, contract No. W-31-109-ENG-38 (JDJ and DGH). \ Useful discussions
with A.-K. Klehe are acknowledged.

\newpage

\begin{center}
{\Large FIGURE \ CAPTIONS}
\end{center}

\bigskip \ 

\noindent {\bf Fig. 1. } Real part of the ac-susceptibility versus
temperature for overdoped Hg-1201 (sample A) at ambient pressure. \ Data
points become very dense above 40 K. \ Solid line is guide to the eye. \ The
inset shows a blow-up of the data near the superconducting onset temperature 
$T_{c}$.\bigskip

\noindent {\bf Fig. 2. \ }Dependence of $T_{c}$ on hydrostatic pressure for
two samples of overdoped Hg-1201: (a) $T_{c}($1 bar) $\simeq $ 47.8 K
(sample A); pressure was successively increased at 298 K (o) and then
successively decreased at either 298 K ($\blacksquare $) or at temperatures 
\mbox{$<$}%
55 K ($\blacklozenge $). \ (b) $T_{c}($1 bar) $\simeq $ 40.0 K (sample B);
pressure was successively increased at 298 K (o) and then successively
decreased at temperatures 
\mbox{$<$}%
65 K ($\bullet $), with numbers giving the order of measurement. \ Pressure
was then reapplied ($\bigtriangleup $) at 298 K and released ($%
\blacktriangle $) at T 
\mbox{$<$}%
65 K, as before. \ See text for full details. \ Solid lines are guides to
the eye.\bigskip

\noindent {\bf Fig. 3. } Dependence of $T_{c}$ on annealing temperature T$%
_{a}$ in overdoped Hg-1201 (sample A) following the application of pressure
at 298 K and its removal at 50 K. \ Relaxation effects are only observed for
temperatures above 200 K. Numbers give order of measurement. \ Dashed and
solid lines are guides to the eye.\bigskip

\noindent {\bf Fig. 4. } Relaxation data on overdoped Hg-1201 (sample A) at
three different pressures. \ The solid lines give fits using Eq. (1) with
given values of $\tau $. \ Numbers give order of measurement. \ See text for
details.\bigskip

\noindent F{\bf ig. 5. }Dependence of $T_{c}$ on pressure in underdoped
Hg-1201 where 0.7 GPa pressure is applied at 298 K but released at
temperatures below 90 K. Numbers give order of measurement. \ Dashed and
solid lines are guides to the eye. \ See text for details.\bigskip

\noindent {\bf Fig. 6. } Activation energy versus pressure for overdoped
Hg-1201 (sample A). \ Dashed line gives best linear fit to data.


\begin{references}
\bibitem{comment-density}  For this estimate the x-ray density $\rho $ =7
g/cm$^{3}$ is used; the actual density of the polycrystalline sample is
roughly 60\% of the X-ray density.
\bibitem{comment-on-alpha}  A detailed analysis of the optimal value of $%
\alpha ,$ as carried out for Y-123 (See Ref. 31) is not possible here
because of the small shifts in $T_{c}$. Nevertheless, $\alpha $ = 0.6
consistently gives the best fits for Hg-1201.
\bibitem{note1}  In a simple Gr\"{u}neisen model, $d\ln \tau _{o}/dP\simeq
\kappa (d\ln \tau _{o}/d\ln V)\approx -\kappa \gamma \approx $ -2\%/GPa,
where $\kappa \approx $ 1\%/GPa is the compressibility and $\gamma \approx $
2 is the Gr\"{u}neisen constant. \ This change in $\tau _{o}$ is negligible
compared to the measured change in $\tau (P)$ itself, where $d\ln \tau
/dP\simeq $ +500\%/GPa.
\end{references}
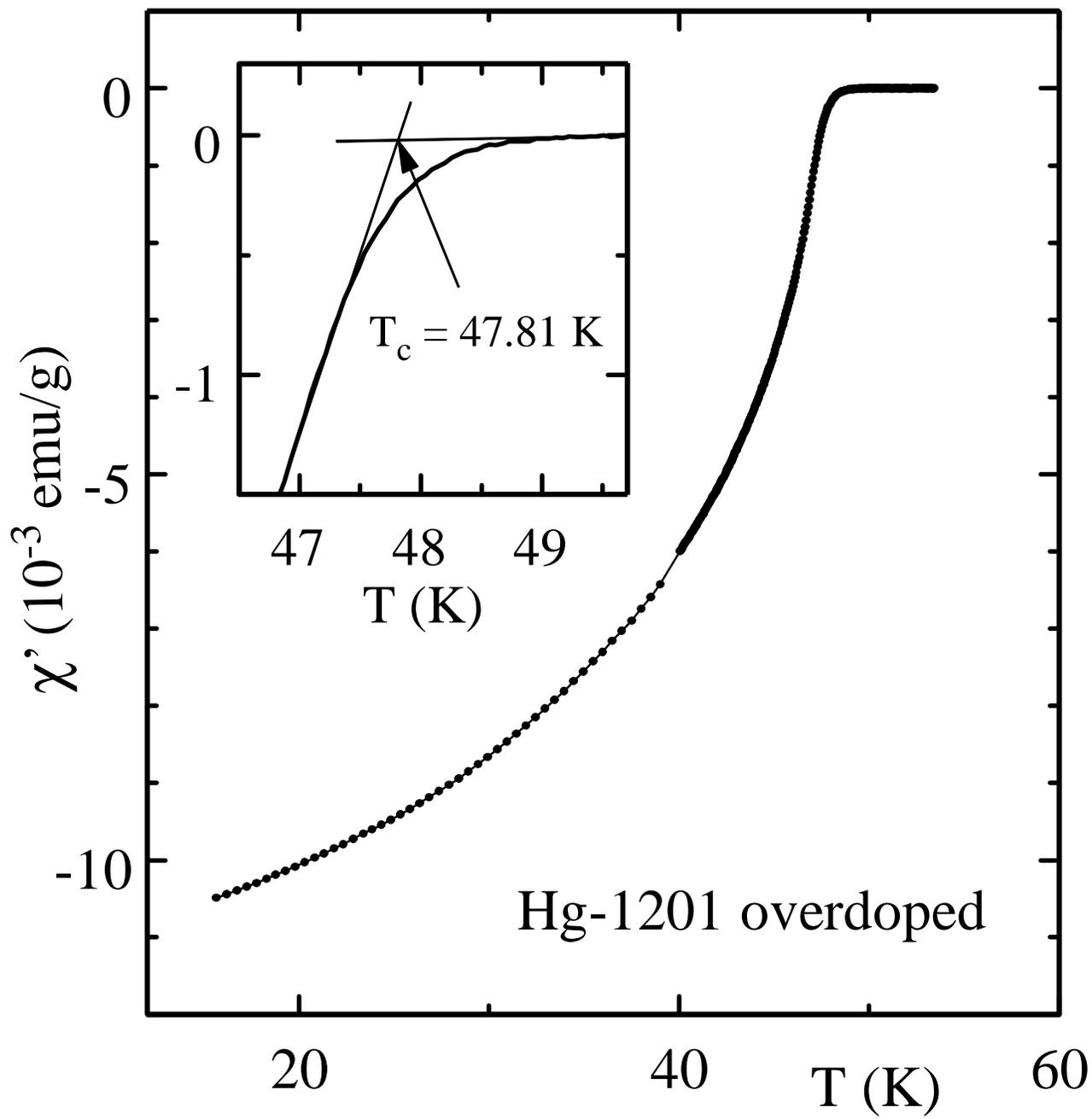
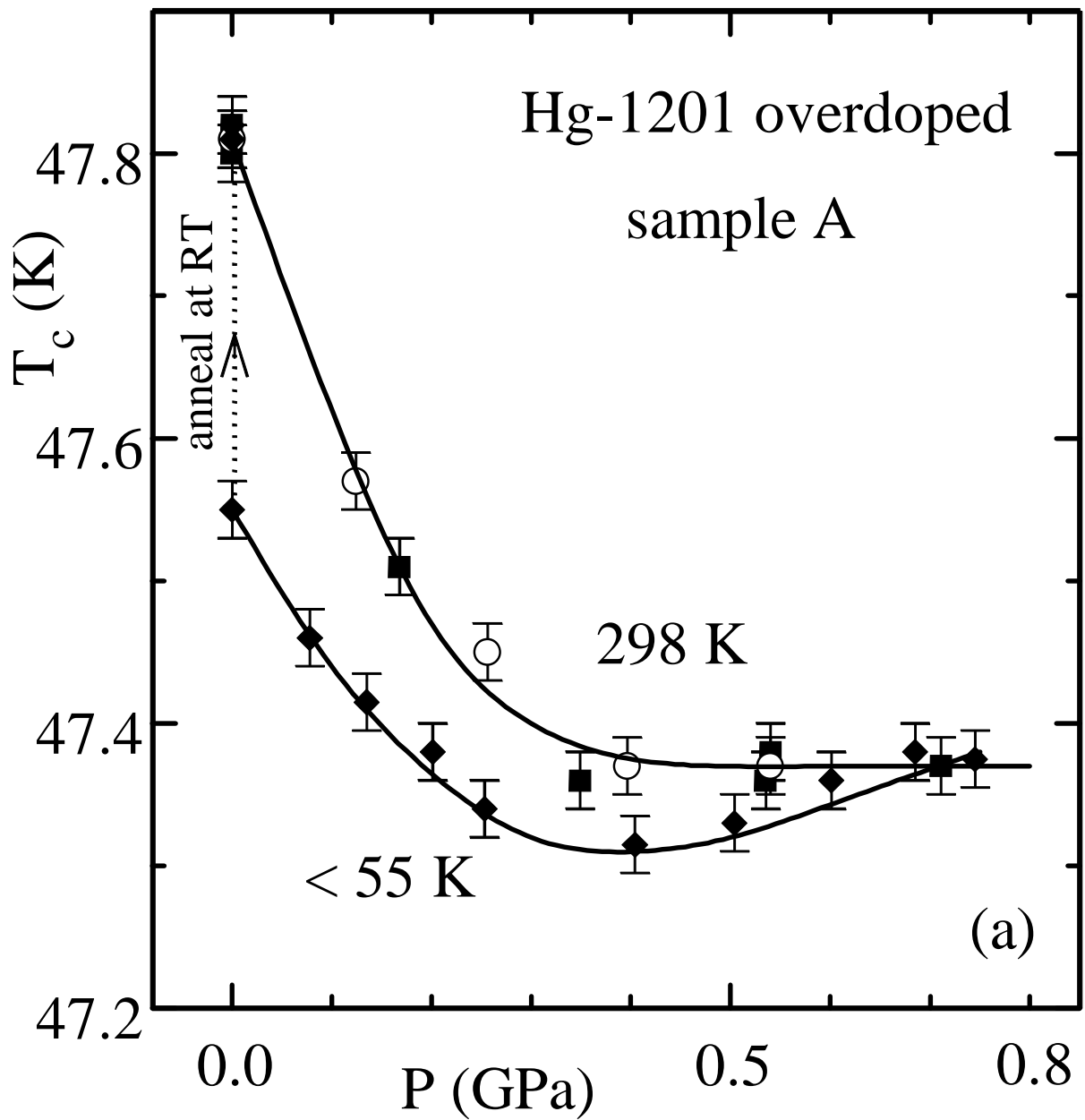
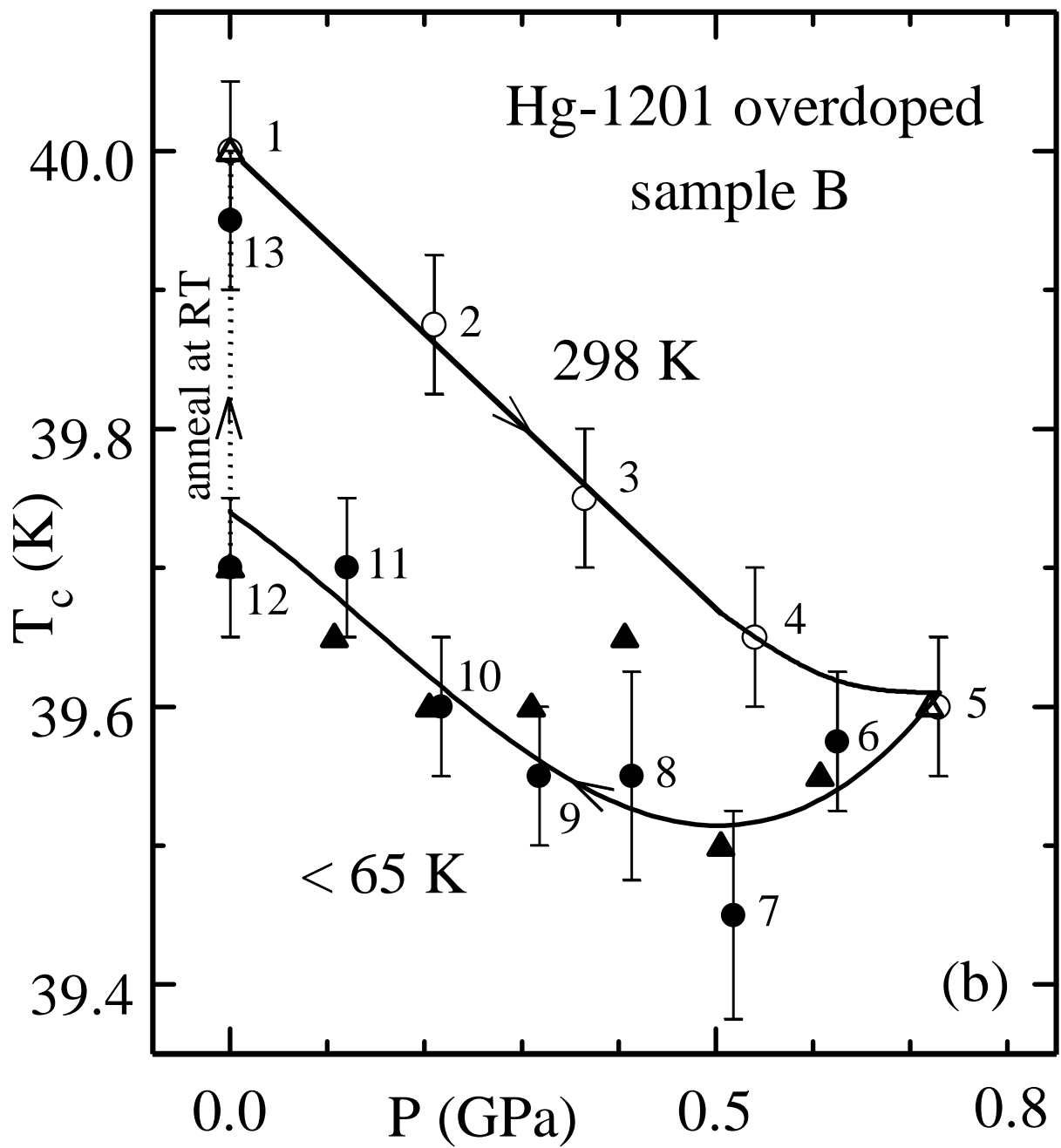
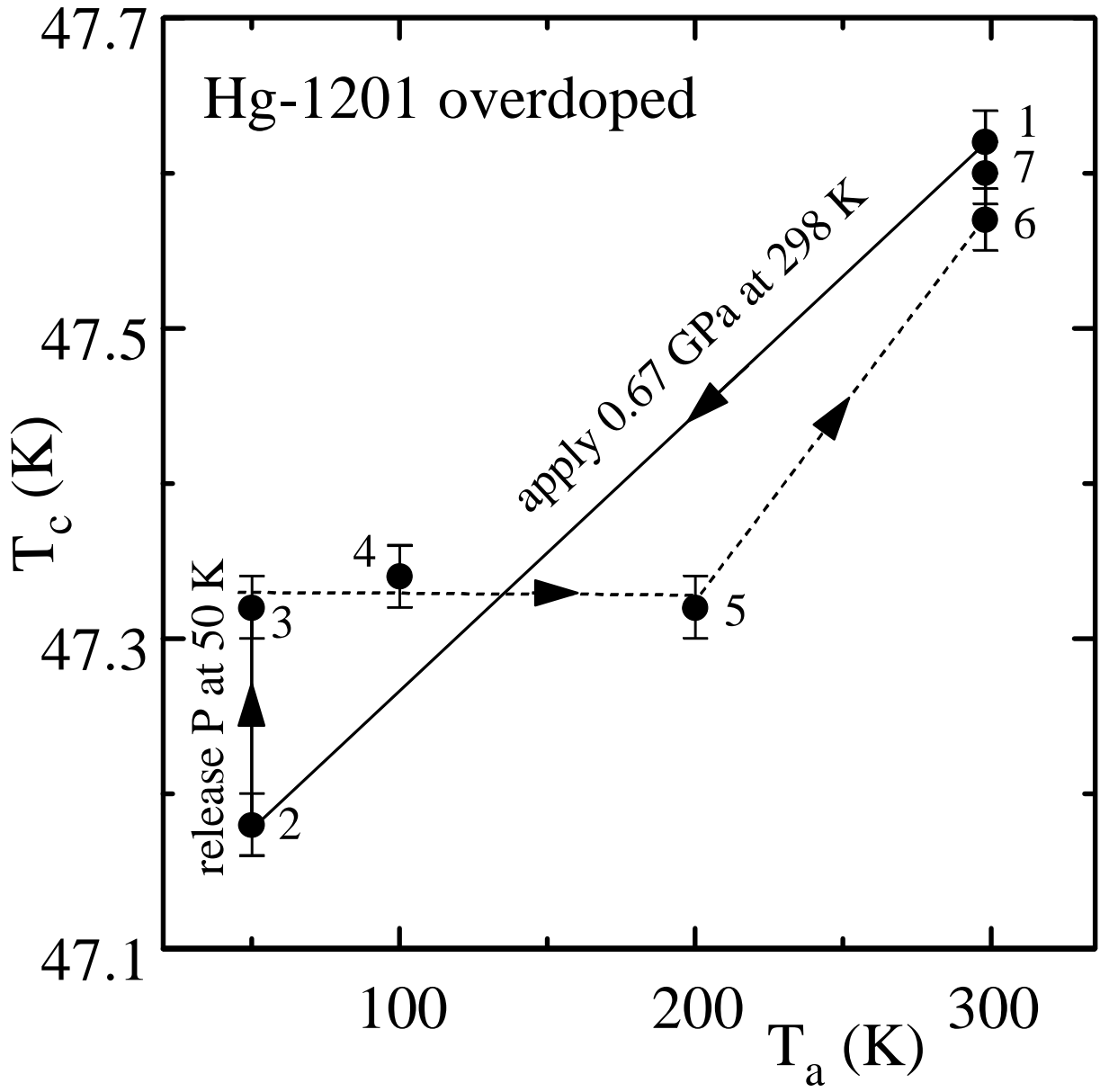
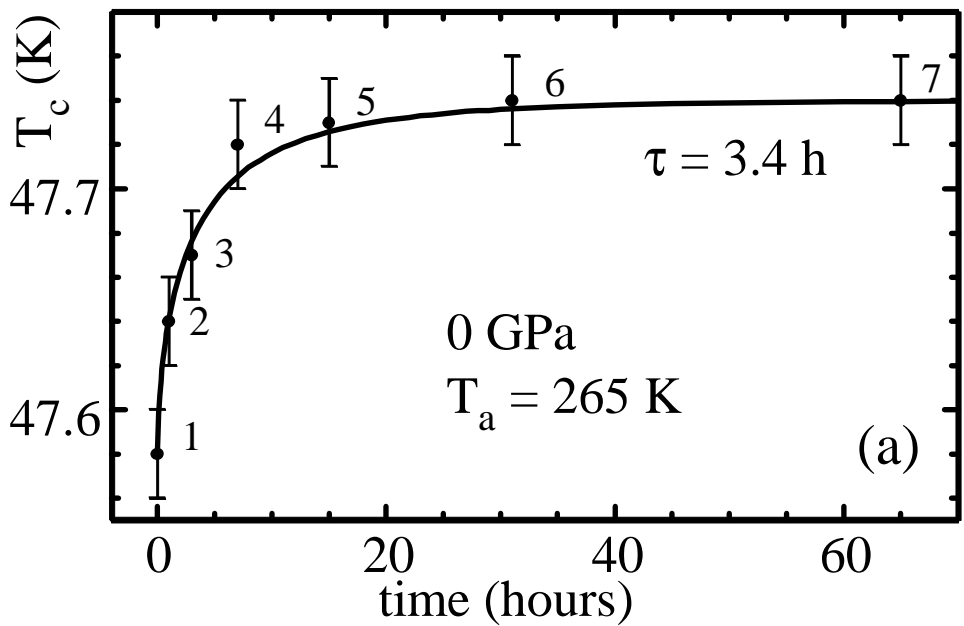
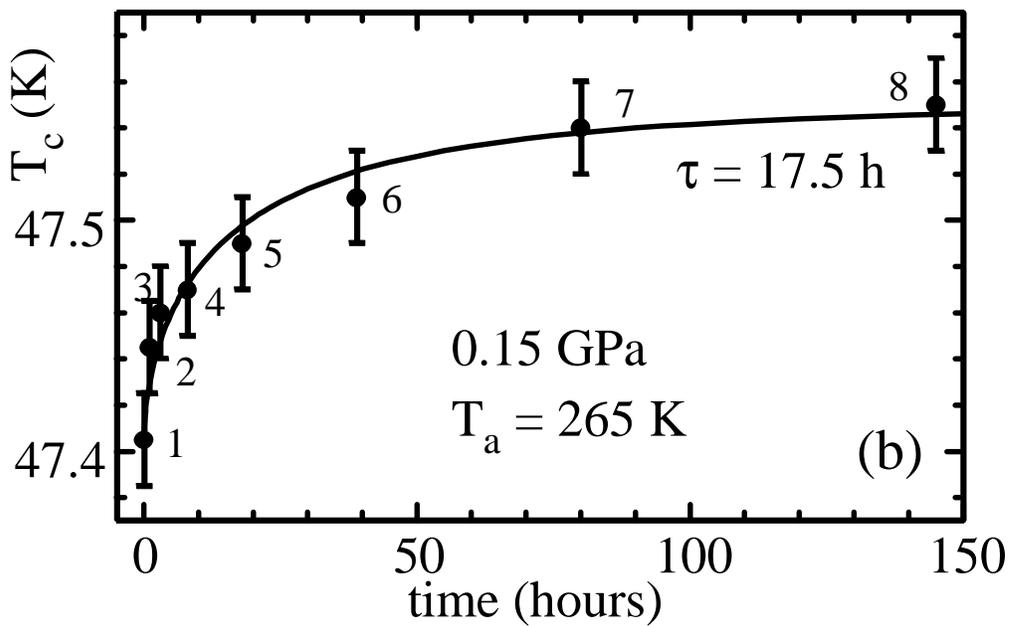
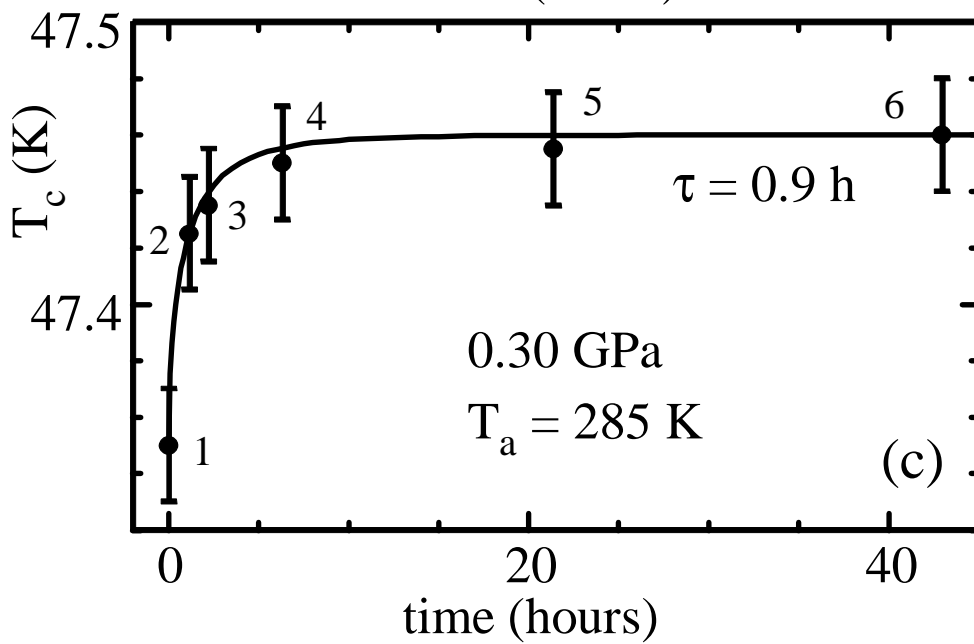
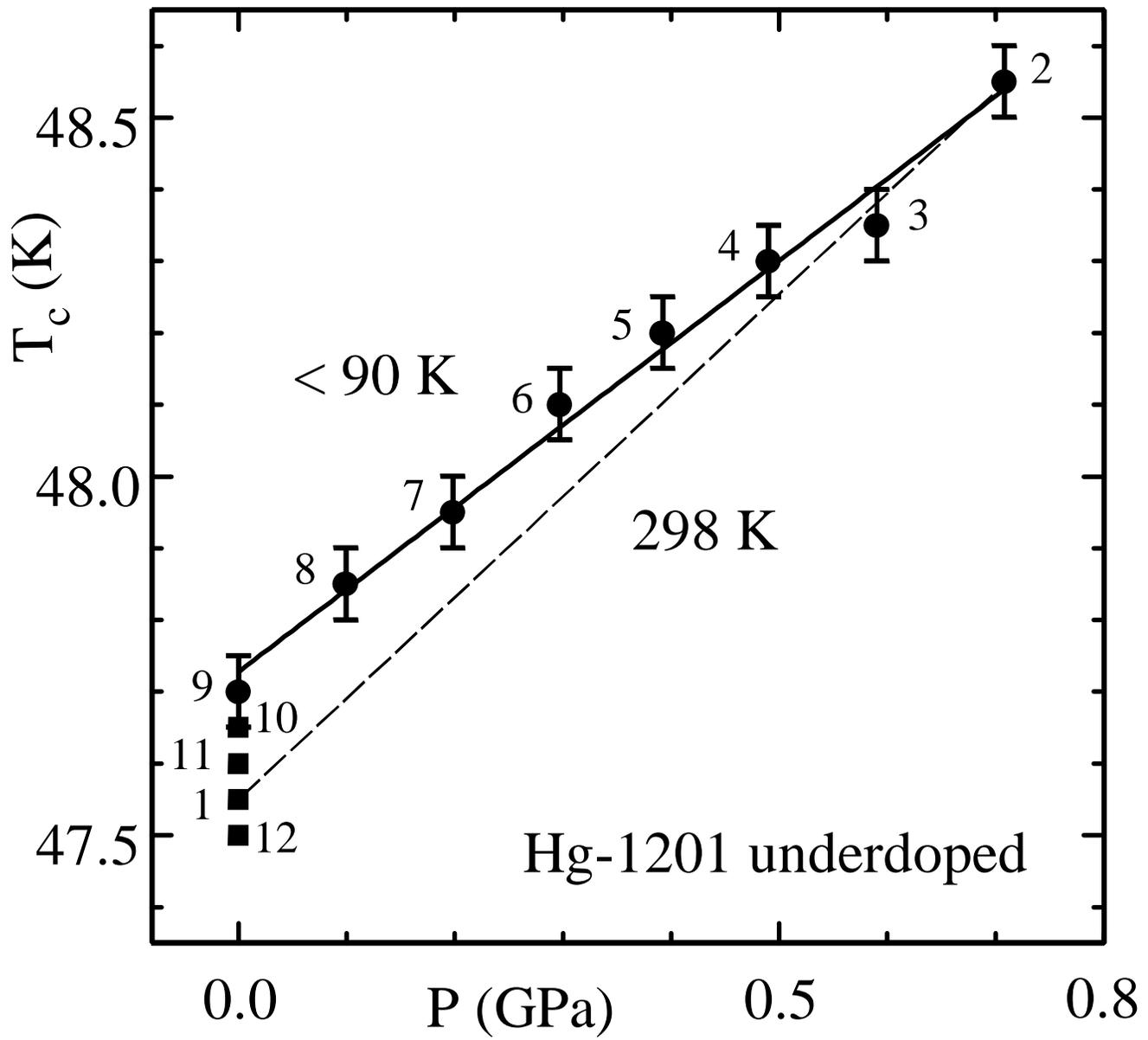
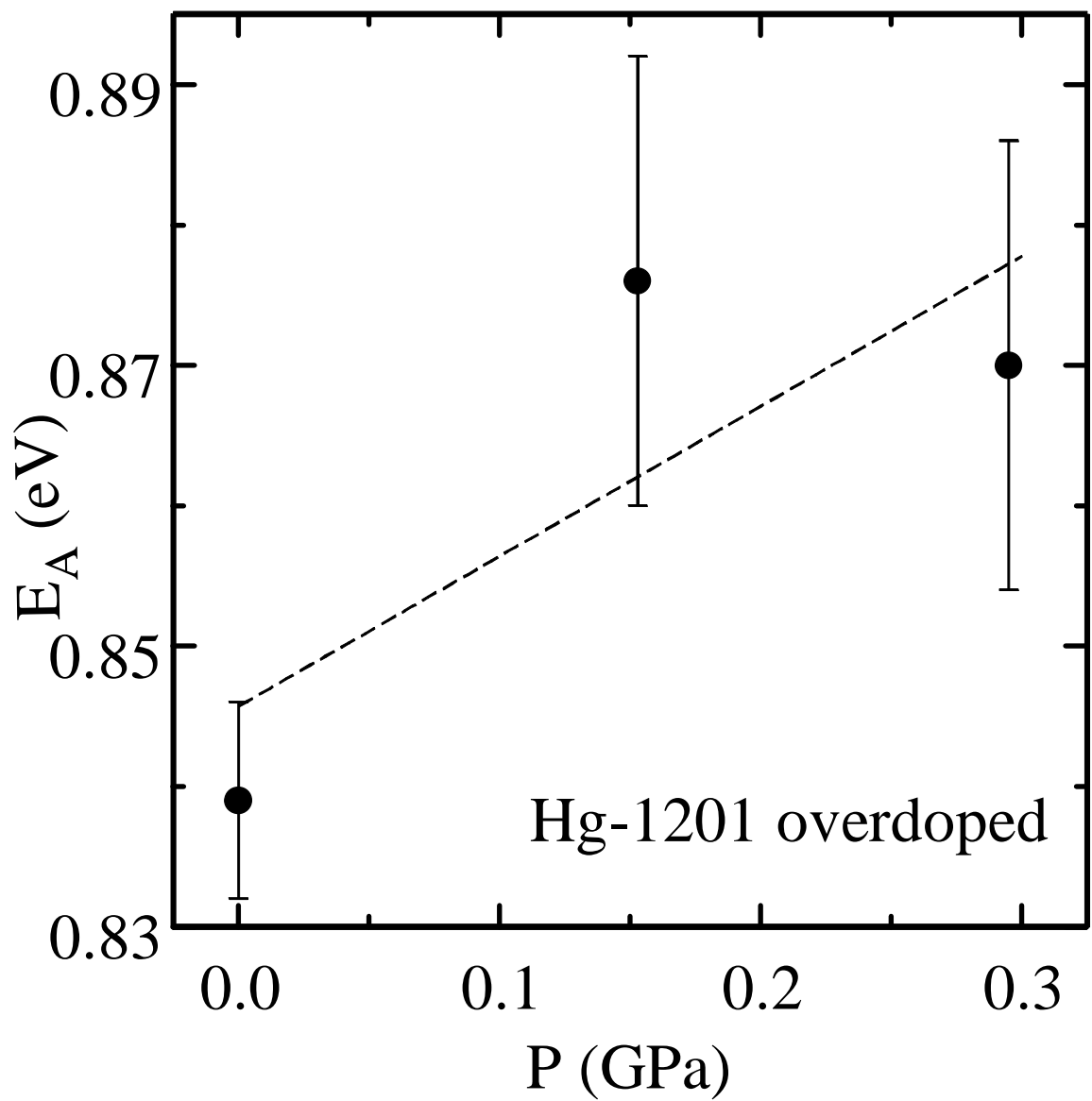
\end{document}